\documentclass[epjST]{svjour}
\usepackage{graphics}
\usepackage[colorlinks,citecolor=blue,linkcolor=blue,urlcolor=blue]{hyperref}
\renewcommand{\phi}{\varphi}
\renewcommand{\>}{\right \rangle}
\newcommand{\<}{\left \langle}
\newcommand{\ket}[1]{\left |#1\>}
\newcommand{\bra}[1]{\<#1\right |}
\newcommand{\be}{\begin{equation}}
\newcommand{\ee}{\end{equation}}
\newcommand{\bea}{\begin{eqnarray}}
\newcommand{\eea}{\end{eqnarray}}

\begin{document}
\title{Multiple Ionization under Strong XUV to X-ray Radiation}
\author{P. Lambropoulos\inst{1,2} \and G. M. Nikolopoulos \inst{1}
}                     
%
%
\institute{Institute of Electronic Structure \& Laser, FORTH, P.O.Box 1385, GR-71110 Heraklion, Greece \and Department of Physics, University of Crete, P.O. Box 2208, GR-71003 Heraklion, Crete, Greece}
\date{Received: date / Revised version: date}
%
\abstract{
We review the main aspects of  multiple photoionization processes in atoms exposed to intense, short wavelength radiation. The main focus is the theoretical framework for the description of such processes as well as the conditions under which direct multiphoton multiple ionization processes
can dominate over the sequential ones. We discuss in detail the  mechanisms available in different wavelength ranges from the infrared to the hard X-rays. The effect of field fluctuations, present at this stage in all SASE free-electron-laser (FEL) facilities, as well as the effect of the interaction volume integration, are also discussed. 
%
} 
\maketitle
\section{Introduction}
\label{intro}
If a system with bound electrons, such as for example atom or molecule, is submitted to electromagnetic (EM) radiation of any frequency, given sufficient intensity and appropriate pulse duration, will eject
several electrons; a fact known for 40 years. The degree of such multiple ionization, by which we mean number of electrons ejected after the pulse is over, does of course depend on the characteristics of the source, as does the mechanism of photoionization. A question that arose, explored and debated at least since the 1980's is whether in cases of multiple ionization, all electrons are ejected sequentially or directly (more than one at the ``same time") \cite{Luk83,BoyRho85,PL85,PL87,Rob86}, i.e. without going through a succession of ionic stages.  Before moving on with our discussion, however, we need to sharpen our terminology, and in particular basic terms such as intense (strong) and short pulse duration.

For photon energies in the infrared (IR) and up to ultraviolet (UV), say up to a few eV, peak intensities
above $10^{13}$W/cm$^2$ and pulse duration sub-picosecond, the dominant mechanism of electron ejection is
tunnelling. To the extent that a second or perhaps third electron may be ejected in that case, by far the
dominant mechanism is what is referred to as recollision \cite{comment,ScrJPB06}. The underlying physical picture rests on
the idea that the first electron is pulled out by the field, undergoing oscillations with the frequency
of the radiation. In that processes, upon returning to the core, it can either recombine giving rise to
HOHG (High Order Harmonic Generation) or be ejected with a spectrum of kinetic energies spaced by the
photon energies, known as ATI (Above Threshold Ionization). Provided the kinetic energy of the returning
electron is larger than the binding, or at least the excitation, energy of a second valence electron, two
electrons may depart, an effect referred to as non-sequential double ionization. 
The reader unfamiliar with the field might observe that a term like ``non-sequential" signifies what the mechanism 
is not, but leaves unanswered the question of what it is. In the community of the IR strong field, the alternative was 
understood to be the so-called recollision mechanism, described immediately below. 
The central theme and objective of this paper is to introduce and discuss in detail an entirely different alternative, 
which emerges under short wavelength strong field, where the recollision option is inapplicable.

The kinetic energy of the
electron undergoing the oscillations under the field is characterized by the ponderomotive energy $U_{p}$,
which corresponds to the cycle-averaged kinetic energy of a free electron in the field. This quantity is
also very close to the AC Stark shift of a highly excited (Rydberg) electron \cite{Hul} and serves as a
convenient measure of the strength of the field. It is well known that non-sequential double ionization
becomes  significant for intensities well into the non-perturbative regime, which means that  $U_{p}$ must
be much larger than the photon energy. It could be said that the intensity at which $U_{p}$ exceeds the
photon energy signifies the threshold for the onset of non-perturbative behaviour, characterized by
substantial appearance of ATI photoelectron energy peaks and the associated HOHG. As a point of
calibration, let us note here that for IR radiation of photon energy 1eV and intensity $10^{13}$ W/cm$^{2}$,
the ponderomotive energy is 1.63 eV,i.e. slightly larger than the photon energy. Therefore an intensity
$10^{14}$ is already high intensity for IR radiation. That is why the intensities commonly employed in
studies of ATI and HOHG, driven by radiation at about 800 nm are in the range of $10^{14-15}$ W/cm$^{2}$.
The pulse duration, to which we come later on, is an equally crucial parameter in this context.

The ponderomotive energy is given by the equation 
\[
U_{p}=I/\omega^{2}, 
\]
with $I$ the intensity and $\omega$ the
photon frequency. The dependence of $U_{p}$ on the photon frequency is of pivotal importance in our
considerations in this paper, as it makes it manifestly clear that, with increasing photon energy, the 
ponderomotive energy decreases rather rapidly. It is equally obvious that the ratio of $U_{p}$ to the 
photon energy, which as noted above represents a key criterion for the onset of tunnelling and non-
perturbative behaviour, increases linearly with intensity, but decreases with the third power of the
frequency. As a consequence, for photon energy 100 eV and intensity $10^{15}$W/cm$^{2}$, using the above
discussed numerical example for photon energy 1 eV, we obtain the value $U_{p}=1.63\times 10^{-2}$eV, which is
four orders of magnitude smaller than the photon energy. It is thus clear that, while an IR intensity of
$10^{15}$W/cm$^{2}$ corresponds to a very strong field, for a soft X-ray of 100 eV, it is a weak field, well below
the onset of non-perturbative behaviour. Another quantity used often as a criterion for the onset of
non-perturbative behaviour is the Keldysh parameter $\gamma$, related to $U_{p}$ through the equation
\[
\gamma = \sqrt{\frac{E_B}{2U_p}},
\]
where $E_B$ is the binding energy of the electron.  
Typically, when $\gamma \ll 1$ tunnelling is expected to be the dominant mechanism, otherwise the interaction is
governed by multiphoton (MP) absorption, described in terms of a generalized cross section of the 
appropriate order. Not accidentally, when the Keldysh parameter is much smaller than one, the 
ponderomotive energy is much larger than the photon energy, and vice versa. The two criteria are therefore
mutually consistent and in the remainder of this paper we will refer only to $U_{p}$. It should be kept in
mind, however, that as usual sharp demarcation lines between regimes of different processes do not exist.
This means that in borderline situations, both tunnelling and MP ionization may play significant roles.

The intensity of a pulse, which by convention refers to the peak intensity, is only half of the story.
It is the combination of pulse duration and peak intensity that determine the nature of the dominant
processes. Clearly, even if at the peak intensity tunnelling is expected to dominate, as the pulse rises
to its peak, the atom experiences  lower but still substantial intensities. For a pulse of long duration,
the atom may loose electrons, one by one, through MP ionization, to such an extent that by the time the peak value is reached, the neutral has been depleted. In fact, if the pulse is too long, several ionic
species may have been produced during the rise of the pulse, so that when the peak value is reached, the
prevalent higher ionic species, being more strongly bound (higher ionization potential), are below the
tunnelling regime. Again, what is long and short pulse is context dependent. It is known by now that non-perturbative behaviour for IR pulses of 780 nm wavelength, requires sub-picosecond pulse durations,
ideally in the range of 10 or so femtoseconds. Otherwise, the population of the neutral is depleted
through a sequence of successive ionization steps, without ever experiencing the peak intensity. To
summarize, for tunnelling and therefore non-perturbative behaviour to be dominant, the ponderomotive
energy corresponding to the peak intensity must be significantly larger than the photon energy, and in
addition the pulse duration must be sufficiently short; in the sense to be further refined below. 

Another very important aspect of strong IR interactions is the so-called Single Active Electron (SAE)
model or approximation. Long wavelength radiation, such as IR and optical, interacts with valence
electrons. No matter how large the intensity, long-wavelength radiation cannot penetrate below the valence
shell, because of screening. For example, in a rare gas such as Xe, with 6 electrons in the 5p shell, not
even the 5s shell, which is immediately below, can be touched by the radiation, until the 5p has been
``opened" through the sequential ejection of one or two electrons. This property implies that, only one
electron is pulled away by the field and  set into the oscillation described above. As a result, the
theory can be cast in terms of the dynamics of one electron driven by the field, while all other electrons
remain practically untouched; hence the term SAE. And it is only through the collision with this returning
electron that the ejection of a second electron can be mediated; hence the term recollision. The 
alternative mechanism, in which two electrons would be ejected through the direct interaction of each of
them with the field, is here inapplicable. But it is the other way around for shorter wavelengths,
as we will see shortly.

Let us consider now the interaction of bound electrons with short wavelength radiation, of photon energy
larger than, say, 30 eV or so, extending past the range of 20 keV; although our main focus will be in the
range from about 40 to 200 eV, for reasons to be discussed a bit later on. In other words, we will be
interested mostly in the XUV to soft X-ray range. As a point of calibration again, note that for photon
energy 50 eV and intensity $10^{15}$ W/cm$^{2}$, the ponderomotive energy is $7\times 10^{-2}$eV, i.e about
three orders of magnitude smaller than the photon energy. For this short wavelength range therefore, and
peak intensities below $10^{18}$ W/cm$^{2}$, the validity of perturbation theory can be taken for granted.
More precisely, it is Lowest no-vanishing Order Perturbation Theory, often referred to as LOPT, for short.
What is implied here is that, if it takes $N$ photons to bridge the ionization potential (binding energy
(BE)) of an electron, the process is describable in terms of a transition probability per unit time
(rate), given by the product of the appropriate (generalized) cross section and the $Nth$ power of the
photon flux; as obtained from the Fermi golden rule. Since we are dealing with a time-dependent flux,
as determined by the features of the particular pulse, the validity of the notion of the cross section
requires that the pulse be sufficiently long. Recall that in the derivation of Fermi's golden rule, for 
a system submitted to a harmonic perturbation, at the end of the procedure the $\lim{t\to\infty}$ is 
taken, giving rise to the delta function, guaranteeing energy conservation between initial and final
state, and the square of the appropriate matrix element of the perturbation, which for ionization amounts
to a cross section. That is why the pulse needs to be ``sufficiently long", which in our context means
about ten  cycles of the field. If the intensity is within the appropriate range, as delineated above, but
the pulse is extremely short, say two field cycles, perturbation theory may still be valid but the notion
of a cross section is at best problematic. In that case, a different approach, such as the numerical
solution of the time dependent Schrondinger equation may be necessary; as is the case for the high
intensity non-perturbative regime.

A very important clarification is called for in this connection. The duration of a pulse in strong-field
physics, in terms of a unit of time such as fs, is not particularly helpful or even meaningful. In the
context of 780-800 nm IR strong fields, it is customary to speak about short or ultrashort pulses of a few
fs. This is entirely reasonable and consistent with our discussion above, because the period of a field in
that wavelength range (photon energy around 1.58 eV) is about 2.6 fs, which means that even 5 fs duration corresponds to slightly less than 2 cycles. On the other hand, at photon energy of 50 eV, 5 fs correspond
to about 63 cycles. Combining our discussion on peak intensity with that on pulse duration, we see that 
while  for IR, $10^{15}$ W/cm$^{2}$ and few fs pulse duration represents a strong, short pulse leading to
non-perturbative behaviour, for XUV and higher photon energies, tunnelling, re-collision, etc. are of no
significance. The electron field interaction in that range of parameters is within the validity of LOPT
with the notion of the cross section perfectly valid. It bears repeating that citing so many Watts per
square cm of intensity and so many fs pulse duration, is totally meaningless, if those numbers are not
examined in the context of the photon energy. Needless to say that, with increasing photon energy, the 
threshold intensity for strong field goes up, and the duration for a short pulse goes down.

There is yet a third important issue to be delineated when it comes to the interaction of XUV and shorter
wavelength radiation with bound electrons. Take, for example, photon energy of 100 eV. Unlike IR or 
optical radiation which interacts with valence electrons, at this higher photon energy, the field has a
stronger coupling with lower shell electrons. Considering again the case of the Xe atom, by far the 
largest photoionization cross section corresponds to the ejection of a 4d and not a valence electron, for 
which the photoionization cross section at that photon energy is more than two orders of magnitude smaller
than that for the 4d. Moreover, the ejection of a lower shell electron is followed by an Auger decay in
which at least one more electron is ejected \cite{richter,richter09,makris09,LamJPB11}. Clearly, this cannot conceivably be described in
terms of a SAE scheme. Let us pause for a moment and ponder the conditions under which for XUV to soft
X-ray radiation, LOPT would not be valid. To be specific, take again 100 eV photon energy. On the basis of
the numbers we derived above, the peak intensity would have to be above $10^{18}$ W/cm$^{2}$ and the pulse
duration less than 100 attoseconds. Unless both of these conditions are satisfied, LOPT will be the valid
tool. Otherwise, we are faced with a daunting problem: The non-perturbative solution of the time-dependent 
Schr\"ondinger equation, without the SAE approximation. Since, as noted above, at this photon energy the
participation of the 4d shell is dominant, with the concomitant Auger decays, it is the dynamics of 
at least 18 electrons under the field that need to be included in the calculation. Obviously, for higher
photon energies, assuming the appropriately higher peak intensity and shorter pulse duration, more
electrons need to be included. As of this writing, this program has been implemented for 2 electrons in
the Helium atom \cite{foumouo,mosch}, as discussed in some detail in the next section.

\section{MP processes under short wavelength radiation: Sequential versus direct processes}

For the time being, the only sources that can muster enough intensity in the XUV range and beyond, are 
the accelerator-based Free Electron Lasers (FEL), the most recent one having provided beams of photon
energy up to 20 keV \cite{J30}. Depending on the machine and the photon energy range, the pulse durations can
be as short as a few fs, but in most cases, they are closer to tens of fs. As established above, the
envelops of those pulses span tens or hundreds of cycles, depending on the magnitude of the central
frequency.
The peak intensities that have been achieved, correspond to ponderomotive energies much smaller than 
the respective photon energies. It is therefore safe to formulate the problem in terms of LOPT with the
appropriate in each case generalized cross sections; or cross sections for short. Since our stated main 
objective is the exploration of the mechanisms leading to multiple ionization, we need a formulation
accounting for the generation and depletion of successive ionic species. It should be obvious that, the
simplest but not the only route towards this end would be the sequential stripping of one electron at a
time, leaving the generated ion in its ground state, and so on. It should be equally obvious that a set
of rate equations could account for the evolution of the ionic species during the pulse, whose solution
at the end of the pulse would provide the final yields for the various ionic species \cite{makris09,LamJPB11}.

The form and underlying physics of such rate equations depend on the photon energy range. For hard X-rays,
the main mechanism is the single-photon ejection of deep inner shell electrons, followed by cascades of
Auger processes, in which several more electrons are ejected \cite{J19,FukuPRL13}. Eventually (during the pulse) ionic
species whose valence ionization potential is larger than the photon energy are generated. In principle,
the process of ionization could continue via two-photon ionization. On the basis of the existing evidence
so far \cite{J19}, it appears that such two-photon processes are too weak to make a significant contribution.
This may be reasonable since the cross section for two-photon ionization decreases rather rapidly with
photon energy in the X-ray range. For the XUV to soft X-ray range (let us say up to 300 eV), on the other
hand, the scenario is quite different. Again, the first few events will be dominated by electron ejection
from lower shells, followed by Auger decays with one or two more electrons escaping, but rather quickly
ionic species for which two-,three-, four-photon, etc. ionization channels become significant are reached 
\cite{makris09,LamJPB11}. In that situation, beyond the initial stages of single-photon events, non-linear MP ionization
processes can take over. It may reasonably be expected that those few-photon non-linear processes will
begin  contributing appreciably at times near the peak of the pulse. And being non-linear processes, as
their rates are proportional to higher powers of the time-dependent flux, their yields will inevitably grow faster. 
It is this range of photon energies upon which we will focus here, because the presence of non-linear
processes opens possibilities for direct multiple ionization, as defined and explained in the following
section.

\section{Direct multiphoton multielectron processes} 
What is the alternative to the sequential stripping as defined above? Obviously, the direct ejection of
more than one electron at a time, mediated by at least one photon absorption from each of the ejected
electrons; without the involvement of either recollision or Auger processes. For this to be energetically
possible, the energy of $N$ photons should be larger than the BE of $N$ electrons, in which case we could have
the direct $N$-photon $N$-electron ionization; without going through the successive ionic stages of the
sequential mechanism. A concrete example with numbers would be helpful at this point. The BE of the two
electrons of He is 79.85 eV, which means that radiation of photon energy between 40 and 80 eV can open a
channel for the direct ejection of both electrons. This is in fact a process that has attracted
considerable attention over more than twelve years now \cite{foumouo,Malegat}. An early precursor 
to this type of process can be found in Ref. \cite{Hul2}. Let us emphasize here
that these direct processes are entirely different from the non-sequential double ionization through
recollision, for IR radiation. No recollision can be involved here, because tunnelling is totally
insignificant and in addition the ponderomotive energy being much smaller than the photon energy, is also
much smaller than the BE of a second electron. Another example discussed recently \cite{Emm}
would be the 3-photon triple ionization of atomic Lithium, energetically possible for photon energies
larger than about 68 eV and smaller than 204 eV. The upper limit mentioned in the above two cases simply
reflects the fact that above that photon energy, single-photon double ionization of He, or single-photon
triple ionization of Li become energetically possible. While below that single-photon multiple ionization
threshold, it is only through the MP processes, direct or sequential, that multiple ionization is
possible. 

It may be worth recalling here that single-photon, two- or three- electron ejection are weak processes, as
they rely on correlation. A single photon interacts with and can induce a transition to only one
electron. As a result, for a second electron to be ejected without the absorption of another photon, the
outgoing electron must somehow transfer to it part of the energy of the photon it has absorbed. This is
possible only for interacting particles bound to the same potential, as is indeed the case for
multielectron atoms. On the other hand, if the energy of $N$ photons is sufficient to eject $N$ electrons,
there is no need for electron-electron interaction, i.e. correlation. Moreover, such an $N$-electron
ejection would be possible even for non-interacting particles; which would be strictly forbidden for a
single-photon process. This property has very important implications to the evaluation of the relevant
transition. Actually, the case of non-interacting particles corresponds to the limiting case of a high $Z$
isoelectronic sequence, where the importance of correlation diminishes rather rapidly with increasing $Z$.
For example, for a He-like ion with $Z=20$, single-photon double ionization would be practically negligible,
while 2-photon double ionization would be smaller than in He only to the extent that the matrix elements
become smaller owing to the tighter binding, remaining totally unaffected by the diminished correlation.

As already mentioned above, 2-photon 2-electron escape in He has received extensive attention, both theoretically
\cite{foumouo} and experimentally \cite{mosch}. Nevertheless much remains to be done in the experimental front, especially
in measuring the 2-photon cross section, the value of which has been the subject of much debate over the
last few years. A thorough summary of the status of this issue and related literature can be found
in the recent paper by Malegat et al. \cite{Malegat}. A calculation addressing 3-photon 3-electron escape in Li has appeared
quite recently \cite{Emm}, which at this point, being the first on this problem, has to be
viewed as semi-quantitative. It has, however, pinpointed a number of pertinent questions, as well as an
outline of the range of source parameters needed for its observation. Clearly the cleanest context for an
$N$-photon $N$-electron escape would be an $N$-electron atom. However, in principle the process would be
operative in any atom with more than $N$ electrons, although its observation may be at least partly
overshadowed by the presence of competing channels, other than the purely sequential which are always
present. Examples illustrating this point can be found in \cite{NikPRL03} for 2-photon 2-electron ejection in Mg, as well as
in \cite{Emm} for 3-photon 3-electron escape in Li.

\section{Definition and calculation of multiphoton ionization cross sections}
\label{sec4}
Let us, before embarking on the quantitative exploration of direct multiphoton multiple ionization,
summarize some basic notions and equations on MP (generalized) cross  sections. Although, usually MP 
ionization implies the ejection of one electron through the absorption of several photons, here we
will extend the notion by including the possibility of the MP escape of several electrons. In the course
of our discussion we will outline our conjecture according to which the magnitude of an $N$-photon 
one-electron escape cross section is of the same order of magnitude as an $N$-photon $N$-electron escape cross
section.

The general expression for an $N$-photon electric-dipole transition amplitude from an initial state $\ket{g}$ to a final state $\ket{f}$, within LOPT, is proportional to 
\[
\sum_{a_{N-1}}\ldots\sum_{a_1}
\frac{\bra{f}\hat{\bf D}\ket{a_{N-1}}\ldots\bra{a_1}\hat{\bf D}\ket{g}}{[E_{a_{N-1}}-E_g-(N-1)\hbar\omega]\ldots(E_{a_1}-E_g-\hbar\omega)},
\]
where $\hat{\bf D}$ is the electric dipole operator, and $E_j$ is the energy of the state $\ket{j}$.
For an $M$-electron atom, in principle, all states entering this expression are $M$-electron states. In reality, certain approximations are involved in calculating a multiphoton transition. Thus for $N$-photon single electron ejection, which may include ATI,  the main contribution comes through the SAE approximation. The terms that would contribute in the matrix elements are schematically 
\[
2p\to\{ns,nd\}\to\{np,nf\}\to\{ns,nd,ng\}\to\ldots,
\]
where $n$ here denotes the principal quantum number, 
with the maximum angular momentum for, say, a 6-photon process being 7. In this case, $\ket{f}$ involves one electron in the continuum. 

Anticipating the detailed multi-electron direct process analysed in depth in the next section, let us use here the example of Neon under 93 eV photons.  Consider,  the case of a direct 6-photon, 6-electron transition from an initial state $2s^22p^6$, the corresponding transition amplitude would have the form
\bea
\sum_{{\bf k}_1}\ldots\sum_{{\bf k}_5}&&
\bra{{\bf k}_6\ldots {\bf k}_1;2s^2}\hat{\bf D}\ket{{\bf k}_5\ldots {\bf k}_1;2p2s^2} \times \nonumber\\
&&\bra{{\bf k}_5,\ldots {\bf k}_1;2p2s^2}\hat{\bf D}\ket{{\bf k}_4\ldots {\bf k}_1;2p^22s^2} \nonumber \times \\ 
&&\ldots \frac{\bra{{\bf k}_1;2p^52s^2}\hat{\bf D}\ket{2p^62s^2}}{\Delta_5\Delta_4\ldots\Delta_1}\nonumber
\eea
where the $\Delta_j$ in the denominator denote the energy differences corresponding to the (virtual) intermediate states of energies in the continuum. 
The intermediate states lie in the continuum, in both, the single as well as the multiple ionization amplitude and the summation (integration) over intermediate states involves a principal value part and a delta function. Although, in principle, the summation over intermediate states contains complete sets of eigenstates of the multielectron system, in practice only certain subsets provide the dominant contribution. For the single electron (ATI) ionization, the dominant contribution comes from single electron matrix elements in the SAE approximation. For the direct multielectron process, it is again single electron matrix elements connecting the continua to electrons in the ground state. For the specific case of Neon, this means that the relevant matrix elements connect the $2p$ to the continuum. The main difference between single and multiple ionization is that the former has more angular momenta available  (up to 7), while the latter involves only $s$ and $d$.  But the successive transitions for the single electron involve matrix elements between virtual states of increasingly higher energy, while for the direct process, all matrix elements of importance connect $2p$ electrons to the continuum and are therefore larger than matrix elements between higher energy states. As a result, the larger number of angular momenta is counterbalanced by the relatively larger magnitude of the radial matrix elements. This is the basis for the above mentioned  conjecture which has in addition been tested quantitatively in  2-photon double ionization of Helium \cite{NakNikPRA}, 
the 3-photon triple ionization of Lithium \cite{Emm}, as well as in the 4-photon double ionization of Carbon \cite{PL87}. 
In general, adopting the above conjecture, one may calculate higher order cross sections
through a procedure of scaling, from the single-photon and two-photon
cross sections, which are usually obtained from numerical calculations \cite{PL87,makris09}.

As to the role of correlation in the direct process, the fact that by
definition the electrons of the initial state connect directly to the final
states shows that such a process is possible without any correlation; it
would in fact be possible even for ``non‐interacting electrons". 
The electrons do of course interact, but the above property of the direct
process suggests that single configuration states would encapsulate
most of the strength of the transition. Correlation would make some
difference, especially in details of photoelectron energy and angular
distributions, but not in the total cross section. In any case,  part of the correlation in the
initial state is anyway included in obtaining the cross section from
scaling, which uses information on the ``size" of the state.

\section{A truly multielectron case}

Given the ongoing development of FEL sources, with reasonably anticipated versatility in levels of
intensity, pulse duration, as well as stochastic properties --- an important aspect discussed below --- it is
tempting to explore a more general question, namely the direct $n$-photon $n$-electron escape for $n$
appreciably larger than three. In order to present a quantitative analysis of such a more general process,
we have chosen to discuss here, in some detail, multiple ionization of Neon with up to 6-photon 6-electron escape,
which has appeared in the literature relatively recently \cite{LamPRA11}. Actually the most general scenario would be
direct $n$-photon $m$-electron escape, where $n > m$, an example of which does in fact occur in Neon, as we will
see shortly. We need to ask for the indulgence of the reader,  for the sudden change of notation in this section, 
in which $N$-photon and $M$-electron are renamed $n$-photon and $m$-electron,  necessitated by the use of $N$ 
for the labelling of ionic species yields.

Whereas linear photon
absorption, be it photoionization or photoexcitation, depends on the average intensity, any non-linear
absorption, for which more than one photon must be absorbed within a ``short" time for its completion,
depends on the intensity correlation function of the relevant order. Thus, for example, 2-photon
ionization is, strictly  speaking, proportional to the 2nd order intensity correlation function and not
simply to the square of the average intensity, while $n$-photon ionization is proportional to the $n$th order
intensity correlation function. An intensity correlation function, reflects the stochastic fluctuations
inherent in any EM source; with the exception of Fourier limited few cycle pulses. As a result of these
fluctuations, the temporal profile of a pulse exhibits spikes whose position within the envelop of the
pulse, as well as their amplitude, are random and depend on the nature of the process that generated
the radiation in the pulse. From that point of view, such fluctuations are not necessarily a ``dirt"
effect, but on the contrary contain substantial information about the character of the source. It is in
fact accurate to say that an $n$-photon process represents an $n$-photon coincidence detector. In plain
physical terms, the atom detects the fashion of photon arrivals, whether and when they arrive one by
one or bunched, in bunches of various numbers.

The study of the stochastic properties of EM radiation and in particular its correlation functions has
occupied center stage in quantum optics over the last half a century, as has the influence of the
fluctuations on laser-atom interactions. Specific models, such as a chaotic (thermal) or a field with only
phase fluctuations, have served as standard models, lending also themselves to analytic solutions. For
example, the effect of intensity fluctuations, also referred to as photon statistics, to 2-photon
absorption represents one of the very early studies \cite{KraOC74} of the connection between field fluctuations and
non-linear processes. For well founded reasons, however, the models of field fluctuations in quantum
optics were based on the assumption on stationary and ergodic processes \cite{Good}. Certainly a chaotic radiation
field, representing a source of bosons in thermal equilibrium at some temperature, is a very realistic if
not exact model for radiation from a lamp. A propos, the $n$th order intensity correlation function of 
chaotic radiation is given by ${n!} \langle F\rangle ^{n}$, where $F$ here denotes the flux,
 while for a pure coherent state it is simply $F^{n}$ \cite{Loud}. 
This obviously implies, counter-intuitive as it may seem, that a chaotic (incoherent!) source is more 
efficient by a factor of ${n!}$ in inducing $n$-photon ionization than a purely coherent source. For an
impressive experimental demonstration of this effect many years ago, in connection with 11-photon
single-electron ionization in Xenon, the reader is referred to \cite{LecPRL74}. And of
course a factor of ${11!}$ means an enhancement of about $10^{7}$.

We discuss now  in some detail the multiple ionization of Neon 
under radiation of photon energy 93 eV ($\approx$ 13.3 nm), with up to 11-photon 8-electron escape, which has appeared in the literature relatively recently \cite{LamPRA11}.  For peak intensities as high as $10^{18}\rm{W}/\rm{cm}^2$, the 
corresponding ponderomotive energy $U_{p}$ is about 10 eV, which is much smaller than both the photon energy 
and the binding energy of any electron in Neon, thus guaranteeing the validity of LOPT as discussed above. 
Moreover, one has sufficiently large number of cycles for a pulse duration of more than 1 fs so that 
the notion of the generalized cross section is valid. Thus, the populations of the ionic species 
produced in the process can be described by the following closed set of rate equations
\bea
\label{rate_eqs1}
\frac{dN_0}{dt} &=& -\sigma_{0,1}^{(1)} F N_0 -\sigma_{0,8}^{(11)} F^{11} N_0\nonumber\\
&&-\sum_{n=2}^{6}\sigma_{0,n}^{(n)} F^{n} N_0-\sigma_{0,7}^{(8)} F^{8} N_0\\
\frac{dN_1}{dt} &=& \sigma_{0,1}^{(1)} F N_0-\sigma_{1,2}^{(1)} F N_{1}\\
\frac{dN_2}{dt} &=& \sigma_{0,2}^{(2)} F^{2} N_0+\sigma_{1,2}^{(1)} F N_{1}-\sigma_{2,3}^{(1)} F N_2\\
\frac{dN_3}{dt} &=& \sigma_{0,3}^{(3)} F^{3} N_0+\sigma_{2,3}^{(1)} F N_{2}-\sigma_{3,4}^{(2)} F^2 N_3\\
\frac{dN_4}{dt} &=& \sigma_{0,4}^{(4)} F^{4} N_0+\sigma_{3,4}^{(2)} F^2 N_{3}-\sigma_{4,5}^{(2)} F^2 N_4\\
\frac{dN_5}{dt} &=& \sigma_{0,5}^{(5)} F^{5} N_0+\sigma_{4,5}^{(2)} F^2 N_{4}-\sigma_{5,6}^{(2)} F^2 N_5\\
\frac{dN_6}{dt} &=& \sigma_{0,6}^{(6)} F^{6} N_0+\sigma_{5,6}^{(2)} F^2 N_{5}-\sigma_{6,7}^{(3)} F^3 N_6\\
\frac{dN_7}{dt} &=& \sigma_{0,7}^{(8)} F^{8} N_0+\sigma_{6,7}^{(3)} F^3 N_{6}-\sigma_{7,8}^{(3)} F^3 N_7\\
\frac{dN_8}{dt} &=& \sigma_{0,8}^{(11)} F^{11} N_0+\sigma_{7,8}^{(3)} F^3 N_{7},
\label{rate_eqs8}
\eea
where $N_j$ is the yield for  the $j$th 
ionic species of charge $(+j)$, with the term $\sigma_{j,k}^{(n)} F^{n} N_{j}$ representing 
an $n$-photon process leading from species $j$ to species $k$.  
The corresponding $n$-photon (generalized) cross section is  $\sigma_{j,k}^{(n)}$ while 
the time-dependent photon flux $F(t)$ is given in photons per cm$^2$ per second. 
In general, these equations have to be solved under a certain pulse profile $F(t)$ 
that captures all the essential features of the pulses (i.e., duration, shape, fluctuations, etc). 

\begin{figure}
\centerline{
\resizebox{0.75\textwidth}{!}{%
\includegraphics{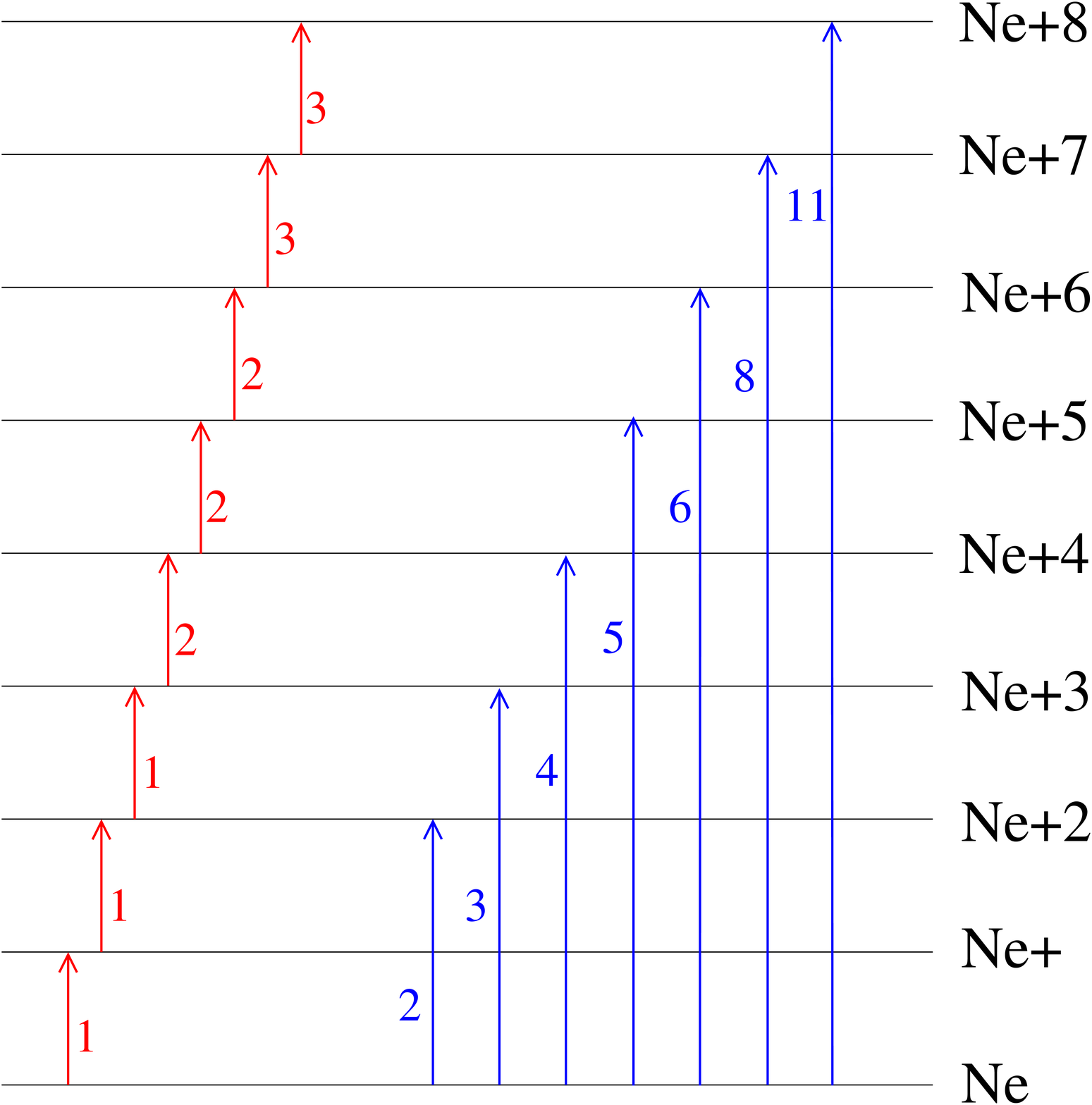}
}}
\caption{Ionization paths of Ne under radiation of photon energy 93 eV ($\approx$ 13.3 nm). The red and blue arrows denote sequential and direct  ionization channels respectively, with the number of photons involved also shown.}
\label{fig:1}       
\end{figure}  
   
   \begin{figure}[htp]
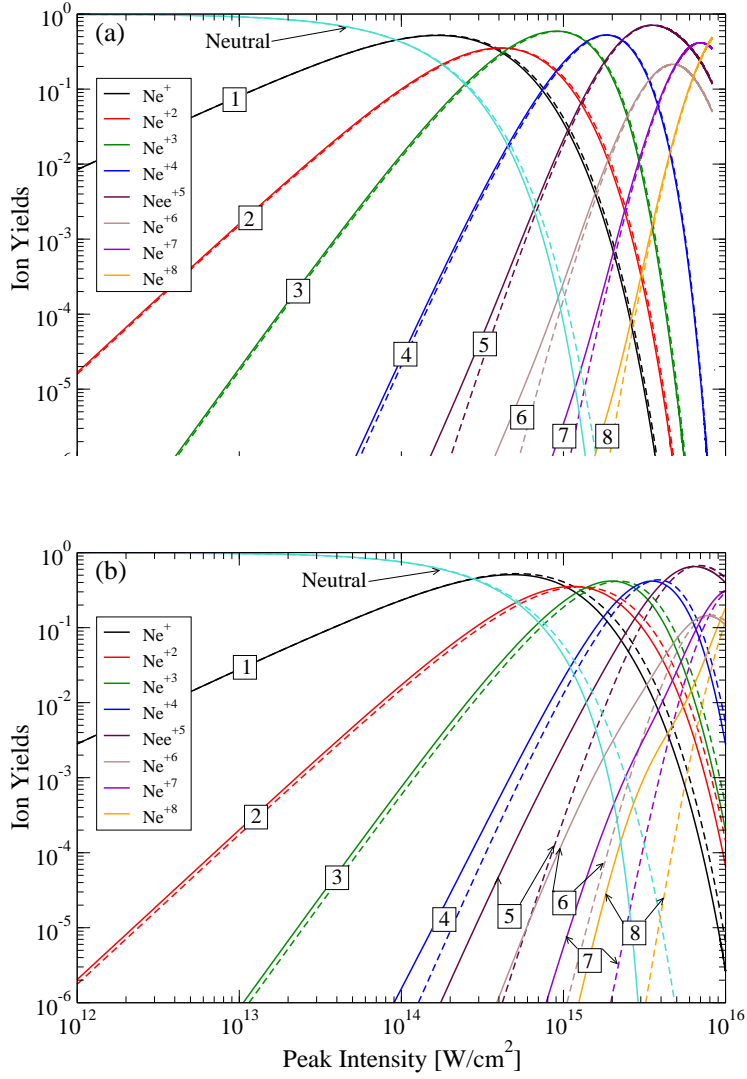

\centerline{
\resizebox{0.75\textwidth}{!}{%
\includegraphics{Deterministic30fs_NewCS-2_Review.eps}
}}
\centerline{
\resizebox{0.75\textwidth}{!}{%
\includegraphics{Deterministic10fs_NewCS-2_Review.eps}
}}
\caption{ Ionization of Ne at 93 eV under fourier-limited pulses in the presence of sequential channels alone (dashed lines)  
and with both direct and sequential processes included (solid lines). The ion yields are plotted as  functions of 
the peak intensity for two different pulse durations: (a) 30 fs; (b) 10 fs. 
} \label{plot1.fig}
\end{figure}

All  ionization paths in the aforementioned rate equations, are summarized schematically 
in Fig. \ref{fig:1}. The sequential processes (red arrows) are associated with single electron 
ejections, and formally can be described by retaining only the terms that lead from ion $j$ to $j+1$ 
in Eqs. (\ref{rate_eqs1}) - (\ref{rate_eqs8}). As discussed earlier, these are expected to be the dominant 
ionization paths  for pulses of large duration and relatively low peak intensity.  On the contrary, for 
short pulses and relatively high peak intensities, a new class of channels 
are energetically possible, which  pertain to multielectron multiphoton
(of the appropriate order) processes, leading directly from the neutral to the
corresponding ions. As shown in Fig. \ref{fig:1} for the system under consideration, 
2 photons can eject 2 electrons leading to
Ne$^{+2}$, 3 photons can eject 3 electrons leading to Ne$^{+3}$, etc., up to 6
photons leading directly to Ne$^{+6}$. These are higher order generalizations
of 2-photon 2-electron ejection in He \cite{foumouo}. 
Moreover, we have included 
an 8-photon 7-electron transition Ne$\to$ Ne$^{+7}$, and an 11-photon 8-electron
transition Ne$\to$ Ne$^{+8}$. Direct $n$-photon $m$-electron
ejection can in principle always occur, for $n\geq m$, as long as it is
energetically allowed. Note that in the above rate equations all  direct ionization channels originate 
from the neutral, neglecting similar channels that originate from intermediate ionic species.  Such processes can be 
neglected since, as will be seen later on, the intermediate ions are drained rather quickly through the 
sequential ionizations channels, and thus they never accumulate a significant amount of population, 
for sufficiently long period of time within the pulse duration. On the contrary, all  population 
at $t=0$ resides at the neutral, and thus direct channels originating from the neutral are expected 
to play the most important role.  In Eqs.  (\ref{rate_eqs1}) - (\ref{rate_eqs8}), these 
processes are represented by the terms on the right hand side which contain the population $N_0$ of the neutral, 
and the multiphoton cross sections of the appropriate order.

Based on the arguments of Sec. \ref{sec4},  the cross sections $\sigma_{i,j}^{(n)}$ employed in our calculations are (in units of cm$^{2n}$sec$^{n-1}$):
\begin{itemize}
\item Sequential processes:
\bea
&&\sigma_{0,1}^{(1)} = 3.52\times10^{-18},\quad\sigma_{1,2}^{(1)} = 1.55\times10^{-18},\nonumber\\
&&\sigma_{2,3}^{(1)} = 1.40\times10^{-18},\quad\sigma_{3,4}^{(2)} = 7.00\times10^{-51},\nonumber\\
&&\sigma_{4,5}^{(2)} = 2.20\times10^{-51},\quad\sigma_{5,6}^{(2)} = 3.00\times10^{-52},\nonumber\\
&&\sigma_{6,7}^{(3)}= 3.00\times10^{-84},\quad\sigma_{7,8}^{(3)}= 5.00\times10^{-85}\nonumber
\eea
\item Direct processes:
\bea
&&\sigma_{0,2}^{(2)} = 7.00\times10^{-51},\nonumber\\ 
&&\sigma_{0,3}^{(3)} = 10^{-83},\quad \sigma_{0,4}^{(4)} = 10^{-115}, \nonumber\\
&&\sigma_{0,5}^{(5)} = 10^{-147},\quad\sigma_{0,6}^{(6)}=10^{-180},\nonumber\\
&&\sigma_{0,7}^{(8)} = 10^{-245},\quad \sigma_{0,8}^{(11)} = 10^{-342}.\nonumber
\eea
\end{itemize}
For an atom like Neon, the further issue of the $2s$ electrons, which are coupled to the
$2p$ owing to the relatively small energy separation of the two subshells, 
has also been in part accounted for in the above equations, through the inclusion in our rates of
ionization from the $2s$ subshell.

To discuss the interplay between sequential and direct ionization paths for Neon at 93 eV, in Fig. \ref{plot1.fig} we 
present the ion yields as a function of peak intensity, at the end of fourier-limited  
pulse of duration 30 fs [Fig. \ref{plot1.fig}(a)] and 10 fs [Fig. \ref{plot1.fig}(b)].
These figures depict the typical single atom behavior, illustrating the appearance 
and disappearance of ionic species as they give rise to higher ones with rising intensity. Of course, as will be discussed later on, 
in practise one has always volume expansion effects that change somewhat the behaviour of the yields shown in Fig. \ref{plot1.fig}. 
In both figures, the dashed lines represent the yields when only the sequential channels are present, whereas the solid curves refer to 
the case of both sequential and direct ionization channels present. In the case of 30 fs  pulses, the curves differ slightly for rather high 
peak intensities. This was to be expected since as mentioned above, the  contribution of direct ionization channels is expected to 
be more pronounced for short pulses, where sequential channels do not have as much of a chance to drain the neutral species by the time 
the pulse has reached its peak. At the same time, the neutral has to be exposed to sufficiently high peak intensities, since the direct 
channels are of higher nonlinearity. Indeed, as depicted in Fig. \ref{plot1.fig}(b), reducing the pulse duration to 10 fs, 
the curves for sequential and sequential+direct ionization channels deviate considerably for ion species above Ne$^{+3}$. 
In contrast, the differences of the curves up to Ne$^{+3}$ are not so prominent because the sequential and the direct channels 
in this case are more or less of the same order. 

The discussion so far has been restricted to the case of fourier-limited pulses, and the main conclusion is that for the enhancement of 
direct channels one has to shorten considerably the duration of the pulse. Current FEL sources, however, produce pulses with strong intensity 
fluctuations (spikes). It has been shown both theoretically and experimentally \cite{SalSchYur,Kri06,Ack07}, that when the FEL operates in the linear regime, the pulses 
exhibit the main statistical properties of polarized chaotic light, with the width of the spikes in the time domain associated with  the coherence time of the light. In contrast to chaotic light typically discussed in quantum optics textbooks, however, FEL pulses cannot be considered either ergodic 
or stationary. Hence, any observable quantities of interest have to be considered in the framework of statistical ensembles over many 
fluctuating pulses.  Experimentally this is achieved by collecting data e.g., for the ion yields, from the interaction of the atomic target with 
many pulses. From the theory point of view, one has to develop numerical algorithms that generate independent randomly fluctuating pulses 
for given coherence time (or bandwidth), nominal duration and peak intensity (the latter two defined with respect to the average pulse). 
Various algorithms have been developed to this end over the last few years and have been applied in various theoretical investigations related to 
experiments \cite{LamPRA11,NikLamPRA12,RohPRA08,PfeiferOL10,NikLamJPB13}. Having such an algorithm available, for each generated random pulse in general one has to solve the equations describing the interaction of the atomic system at hand with the radiation, a procedure typically referred to as realization (or trajectory). At the end, the quantities of interest are averaged over many pulses (realizations).  In this framework, for the example of Neon under consideration, Eqs. (\ref{rate_eqs1}) - (\ref{rate_eqs8}) become stochastic, owing to the stochastic nature of the intensity. 

It is instructive to consider first any one of the direct channels that leads from neutral Ne to a higher ionic species, say Ne$^{+m}$ via an $n$-photon absorption, discarding for the moment any additional channels. 
We thus have only two rate equations namely 
\bea
\label{ex1}
&&\frac{dN_0}{dt} = -\sigma_{0,m}^{(n)} F^n N_0,\\
&&\frac{dN_m}{dt} = \sigma_{0,m}^{(n)} F^n N_0,
\label{ex2}
\eea

Formal integration of Eq. (\ref{ex2}) yields,
\bea
N_m &=& \sigma_{0,m}^{(n)}\int_0^\infty F(t^\prime)^n N_0(t^\prime)dt^\prime\\
&=& \frac{\sigma_{0,m}^{(n)}}{(\hbar\omega)^n}\int_0^\infty I(t^\prime)^n N_0(t^\prime) dt^\prime,
\eea
where $I(t)$ is the intensity and $\hbar\omega$ the photon energy.
For small intensities, 
\[
N_m \approx N_0(0) \frac{\sigma_{0,m}^{(n)}}{(\hbar\omega)^n}\int_0^\infty I(t^\prime)^n dt^\prime
\]
and averaging over many random pulses we obtain 
\bea
\langle N_m \rangle  \approx N_0(0) \frac{\sigma_{0,m}^{(n)}}{(\hbar\omega)^n} 
\int_0^\infty \left \{ \int I(t^\prime)^n p[I(t^\prime)]dI \right \} dt^\prime  
\eea
where $p[I(t)]$ is the probability of having instantaneous intensity between $I(t)$ and $I(t)+dI$. 
For FEL pulses with chaotic properties, this is an exponential distribution and thus the inner integration  
can be performed immediately yielding  
\bea
\langle N_m \rangle   \approx  n! \left [ N_0(0) \frac{\sigma_{0,m}^{(n)}}{(\hbar\omega)^n} 
\int_0^\infty   \langle I(t^\prime)\rangle^n dt^\prime \right ].
\eea
This derivation shows that  $n$-photon ionization, 
within LOPT, is proportional to the $n$-th order intensity correlation function, 
which for a chaotic field is given by $n! \langle {I}\rangle ^n$. As a result, the average ion yield 
is $n!$ times the ion yield for a fourier-limited pulse of the same average pulse duration and 
peak intensity.  

This procedure would be rigorous if we had a single $n$-photon process and in addition 
the field were truly chaotic. What we have, however, in the case of Neon is a set of differential equations 
with various intermediate ionic species, and different ionization channels of various orders. 
Yet, based on the above observation, one may still 
try to replace $F^n$ by $n! \langle F \rangle^n$ in Eqs. (\ref{rate_eqs1}) - (\ref{rate_eqs8}), which amounts to effectively increasing the 
$n$-photon ionization cross section by a factor of $n!$. It turns out that such an approach 
 amounts to a decorrelation approximation of the form
\[\langle F^n(t) N_j(t)\rangle \approx \langle F^n(t)\rangle \langle N_j(t) \rangle\] 
valid only for small intensities where the  
ionic populations do not change appreciably on the scale of the intensity fluctuations. 

 \begin{figure}[htp]
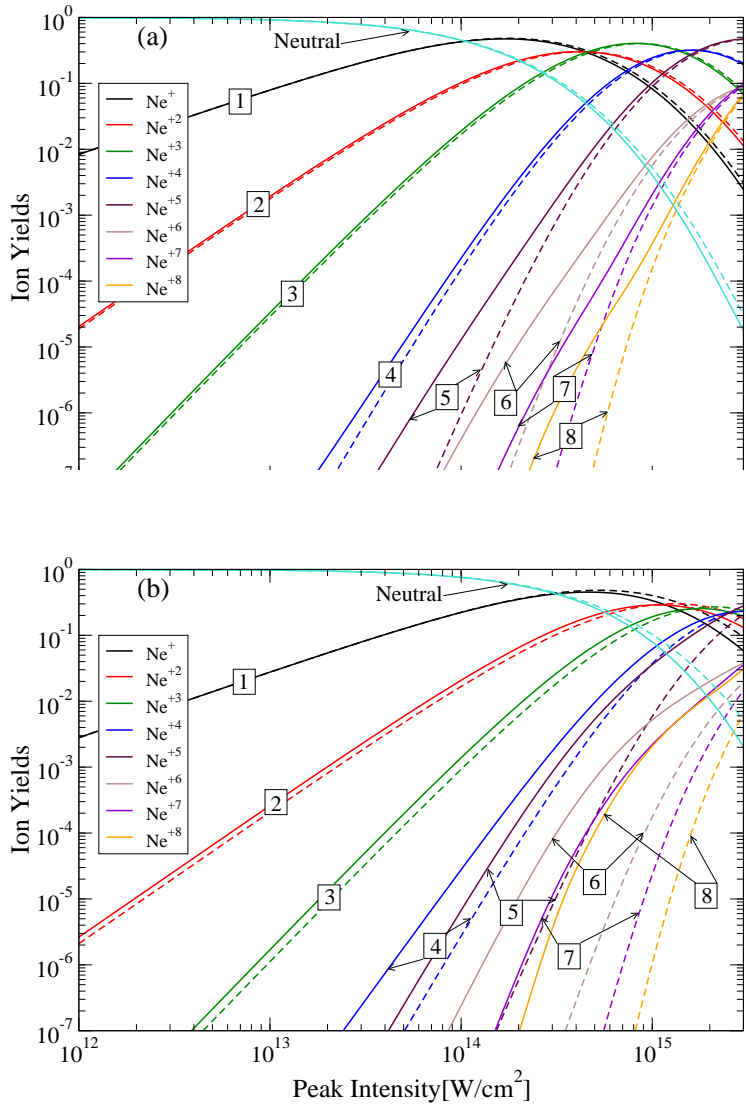

\centerline{
\resizebox{0.75\textwidth}{!}{%
\includegraphics{ChSD_vs_ChS30fs_Tc6fs_10000Trj_NewCS-2_Review.eps}
}}
\centerline{
\resizebox{0.75\textwidth}{!}{%
\includegraphics{ChSD_vs_ChS10fs_Tc2fs_10000Trj_NewCS-2_Review.eps}
}}
\caption{ Ionization of Ne at 93 eV under chaotic pulses in the presence of sequential channels alone (dashed lines)  
and with both direct and sequential processes included (solid lines). The average ion yields are plotted as  functions of 
the average peak intensity for (a) nominal duration 30 fs and coherence time 6 fs; and (b) nominal duration 10 fs and coherence time 2 fs. 
The set of Eqs. (\ref{rate_eqs1})-(\ref{rate_eqs8}) has been solved independently for $10^4$ randomly generated 
pulses $F(t)$. The depicted ion yields have been obtained by averaging over the corresponding single-realization yields.   
} \label{plot2.fig}
\end{figure}

 \begin{figure}[htp]
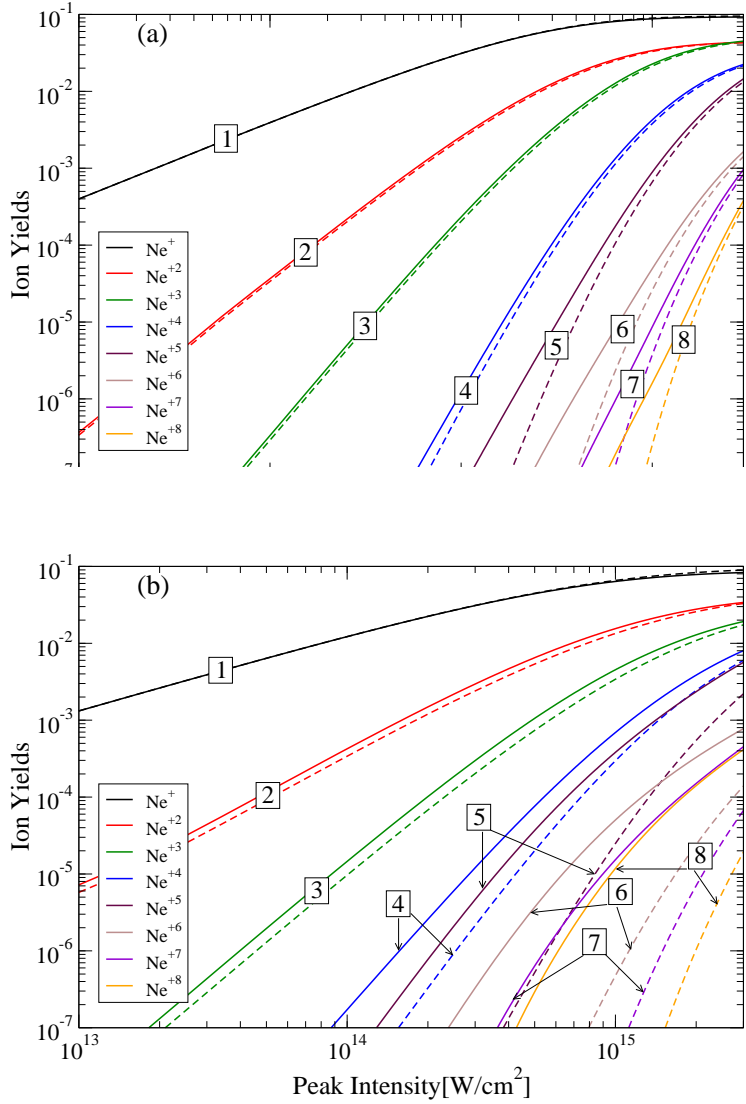

\centerline{
\resizebox{0.75\textwidth}{!}{%
\includegraphics{VolumeIntChSD_vs_ChS30fs_Tc6fs_10000Trj_NewCS-2_Review.eps}
}}
\centerline{
\resizebox{0.75\textwidth}{!}{%
\includegraphics{VolumeIntChSD_vs_ChS10fs_Tc2fs_10000Trj_NewCS-2_Review.eps}
}}
\caption{ As in Fig. \ref{plot2.fig},  with  volume expansion effects included. 
} \label{plot3.fig}
\end{figure}
Hence, the theoretical analysis of the problem for a broad range of parameters can be provided only 
through the above mentioned  rigorous approach namely,   the numerical solution of the  differential equations for 
a sufficiently large number of pulses, and taking the  ensemble average over these realizations. We have performed 
such simulations   for Ne at 93 eV \cite{LamPRA11}, and in Fig. \ref{plot2.fig} we present related results. 
A direct  comparison of the averaged ion yields of  Fig. \ref{plot2.fig} to the results  for fourier-limited 
pulses of the same duration (see Fig. \ref{plot1.fig}) reveals the dramatic effect of fluctuations on the yields of ions 
above Ne$^{+3}$, through  the enhancement of the direct channels. For instance, note that 
for moderate intensities below $10^{15}$ W/cm$^2$
the average yield of Ne$^{+5}$ in the presence of 
direct channels has increased relative to the sequential channels alone by one or even two orders of magnitude (depending on the pulse duration),   whereas the corresponding increases for fourier limited pulses were significantly smaller. 
In general, the individual spikes in a pulse are of shorter duration than the average pulse, and they exhibit higher peaks. 
From another point of view, one can say that a fluctuating pulse is a superposition of many pulses (spikes), of shorter duration and higher 
peak intensities, interpreting thus the enhancement of the direct channels in the context of  fourier-limited pulses. Indeed, the enhancement 
of the direct channels over the sequential ones tends to become more prominent as we decrease (increase) the coherence time (bandwidth) of the light, which is equivalent to adding more narrower spikes of higher peak intensities in a pulse.

To compare with experimental data, one needs to take into account the interaction volume relevant to a particular experiment, and integrate over the spatial distribution of the radiation in this volume.
As shown in Fig. \ref{plot3.fig} the main effect of the volume expansion is to stabilize the populations of the ionic species beyond the saturation intensity i.e., they do not decrease any more,  but rather  exhibit a slow increase, due to the contribution of more atoms from the periphery of the interaction volume, where the intensity is lower than in the center. The crucial point, however, is that the contribution of the direct channels is found to be pronounced, for the higher species, even upon spatial integration. The main message therefore is that  in general  intensity fluctuations in the pulses are capable of enhancing considerably the direct channels over the sequential ones, even for longer pulses, making thus their observation easier in practice, even for setups where volume expansion effects are inevitable. 
 
In view of existing experiments on the ionization of Ne  at 93 eV \cite{richter09}, in Ref. \cite{LamPRA11} we attempted to address the question of whether there are traces of the direct channels in the  reported experimental data, which unfortunately were limited to only peaks of TOF (Time of Flight) results, at a single laser intensity \cite{richter09}. The comparison with the experimental data turned out to be rather  sensitive to the value of the peak  intensity. Nevertheless, our theoretical calculations were  in a  reasonable agreement with the experimental observations for intensities $\sim 3\times 10^{15}\textrm{W/cm}^2$, which suggests that the experimental data have 
been obtained at or beyond the saturation intensity. As is evident in Fig.  \ref{plot3.fig}   these are not the optimal conditions for detecting the contribution of the direct channels.  To this end one needs data over a range of intensities below saturation. 

\section{Concluding remarks}
We began this review by recalling early debates, about 30 years ago, as to whether multiple ionization
observed in the rare gases at the time \cite{Luk83,BoyRho85,PL85,PL87,Rob86} contained the signature of direct or simultaneous escape
of several electrons, in fact of a whole shell. The issue was settled in \cite{PL87} and \cite{Rob86}, where it
was demonstrated that, owing to the combination of photon energies and pulse durations (long pulses), the
resulting yields of multiply ionized species were solely due to the sequential stripping of electrons,
one at a time. The same issue came up and debated anew, in recent years \cite{richter,richter09,makris09,LamJPB11} after the appearance
of FEL's, in experiments under XUV radiation of photon energy around 90-93 eV. It turned out again that,
although this time the photon energies were significantly larger (90 instead of 6 eV), still the 
combination of ponderomotive energy at peak intensity and pulse duration placed the process squarely in
the regime of LOPT, with sequential ionization being the dominant mechanism \cite{LamJPB11}. The accidental
presence of a so-called giant resonance, in the case of Xenon at about 100 eV, did not change the
picture at all. It simply made the first step. which is the single-photon ejection of a lower shell electron more probable \cite{LamJPB11}. The associated Auger processes, being field independent, did not alter the
fundamental sequential mechanism.

Nevertheless we described recently \cite{LamPRA11} and reviewed here a scenario, supported by relevant calculations, in which truly direct, several
electron escape, mediated by an equal or slightly larger number of photons may be observable. The simplest
case of 2-photon 2-electron escape in Helium has already been demonstrated. It remains to be seen whether direct
processes beyond $N=2$, will turn out to be of only academic/theoretical interest, or will indeed be
found to make discernible contributions to the multiple ionization of species under strong XUV to X-ray
radiation. We have shown that  field fluctuations, when present, are not a ``dirt" effect but rather they enrich 
the physical content of the interaction. In any case, until such a time as FEL X-ray pulses become available, the field 
fluctuations need to be taken into consideration. Given the astonishing pace of
development and technological improvement of FEL sources, it is reasonable to expect that the desirable
features, such as shorter pulse durations will sooner or later be realized.   

Assuming that we have Gaussian Fourier limited pulses of sufficiently short duration, say 10 fs, how would we confirm the contribution of the direct channels?  The least demanding approach
would be two sets of data, on ionic yields as a function of laser peak intensity, for two different durations, e.g. 10 and 50 fs. We know that for the longer pulse the yields will be higher. But for
the first and second ionic species, involving linear and only quadratic processes, the increase for the longer pulse will be a simple consequence of the pulse duration easily predictable by the rate equations. 
For the higher ionic species, however,  as we have seen in Fig. \ref{plot1.fig}, the yields for the shorter pulse will be higher, if the direct processes are significant. 
This is an expected consequence of the higher order of non-linearity involved in those channels.
Of course an interplay with theory will be very useful, even if the values of the cross sections are known with modest accuracy, because it is the relative and not the absolute yields for different durations that are to be compared. An experimentally more  demanding manifestation  of the presence of direct contributions should be discernible in the photoelectron energy spectra, because the sequential channels produce isolated photoelectron peaks, while the direct produce continuous distributions \cite{foumouo,Emm,NikPRL03}. It must be kept in mind, however, that significant progress in the FEL sources, in particular in connection with the repetition rate, will be necessary for such an undertaking.

\end{document}